\documentclass[twocolumn,amsmath,amssymb,showkeys]{revtex4}
\usepackage{graphicx}
\usepackage{amsmath,amssymb,mathrsfs}

\def\Tr{\operatorname{Tr}} 
\def\>{\rangle}\def\<{\langle} \def\sH{\mathscr{H}}
   
\def\mE{\mathcal{E}} 
 
 \def\fid{\mathscr{F}}
\def\N#1{\left|\!\left|{#1}\right|\!\right|}

  \def\cohinfo{I_\textrm{c}}
 
  \def\mR{\mathcal{R}}
 \def\eof{E_\textrm{f}}
\def\ccorr{C^{E'\to R}} \def\states{\mathfrak{S}}
\def\id{\operatorname{id}} \def\ode{E_\textrm{d}^{A\to B}}

\newtheorem{theo}{Theorem}
\newtheorem{corollary}{Corollary}

\begin{document}

\title{Entanglement measures and approximate quantum error correction}

\author{Francesco Buscemi} \email{buscemi@qci.jst.go.jp}
\homepage{http://www.qci.jst.go.jp/~buscemi} \affiliation{ERATO-SORST
  Quantum Computation and Information Project, Japan Science and
  Technology Agency,\\ Daini Hongo White Bldg. 201, 5-28-3 Hongo,
  Bunkyo-ku, 113-0033 Tokyo, Japan}

\date{December 20, 2007}


\begin{abstract}
  It is shown that, if the loss of entanglement along a quantum
  channel is sufficiently small, then approximate quantum error
  correction is possible, thereby generalizing what happens for
  coherent information. Explicit bounds are obtained for the
  entanglement of formation and the distillable entanglement, and
  their validity naturally extends to other bipartite entanglement
  measures in between. Robustness of derived criteria is analyzed and
  their tightness compared. Finally, as a byproduct, we prove a
  bound quantifying how large the gap between entanglement of
  formation and distillable entanglement can be for any given finite
  dimensional bipartite system, thus providing a sufficient
  condition for distillability in terms of entanglement of formation.
\end{abstract}

\maketitle

\section{Introduction}

The possibility of performing quantum error correction obviously lies
behind and justifies the vast efforts made up to now in order to
develop quantum computation techniques, since it allows fault-tolerant
computation~\cite{nota1} even when quantum systems---in fact extremely
sensitive to noise---are considered as the basic carriers of
information. Besides well-known algebraic conditions for \emph{exact}
quantum error correction, which directly lead to algebraic quantum
error correcting codes (for a thorough presentation of quantum error
correction theory and a detailed account about the enormous literature
about it, see e.~g.~\cite{nielsen,algebraic-exact}), an
information-theoretical approach to quantum error
correction~\cite{schum-nielsen,ogawa,nielsen-poul} can shed some light
on the dynamical processes which underlie quantum noise, offering at
the same time the opportunity to better understand the conditions
under which \emph{approximate} quantum error correction is
feasible~\cite{schumacher-westmoreland}. In the present paper, we will
be working within the latter scenario.

Approximate quantum error correction is not just a theoretical issue:
in fact, in all practical implementations the experimenter can only
rely upon some confidence level---exact processes exist as abstract
mathematical concepts only. Then, conditions for approximate quantum
error correction can provide useful ways to test the reliability of a
real apparatus. In Ref.~\cite{schumacher-westmoreland}, Schumacher and
Westmoreland proved that an adequate information-theoretical quantity
to consider is the coherent information: the loss of coherent
information along a quantum noisy channel is small if and only if the
quantum noisy channel can be approximately corrected. In a subsequent
paper~\cite{schumacher-westmoreland-2}, the same Authors provided
another criterion, this time for exact quantum error correction: the
loss of entanglement (of formation) is null if and only if the channel
can be exactly corrected. They left open the question whether the loss
of entanglement provides not only a condition for exact correction,
but also a condition for approximate correction. In this paper we will
show that this is actually the case, extending our analysis to
different entanglement measures, thereby proving that many
inequivalent ways to quantify entanglement lead in fact to analogous
conditions for approximate quantum error correction. We will moreover
obtain, as a byproduct, an inequality directly relating the
entanglement of formation with the distillable entanglement present in
a general bipartite mixed quantum state. Such inequality makes
rigorous the intuition, that the gap between entanglement of formation
and distillable entanglement, which is known to exist generically
large for general mixed quantum states~\cite{generic_ent}, cannot be
\emph{completely} arbitrary, in the sense that, given a finite
dimensional bipartite state, whenever the entanglement of formation is
``sufficiently close'' to its maximum value, then also the distillable
entanglement has to be ``correspondingly large''. (The concepts of
``sufficiently close'' and ``correspondingly large'', clearly
depending on the dimensions of the subsystems, will be quantitatively
defined below.)

The paper is organized as follows. In Section~\ref{sec:2} we recall
some basic notions about quantum channels and their purification into
the unitary evolution of a larger closed system. In
Section~\ref{sec:3} we present some known information-theoretical
conditions for exact as well as approximate quantum error
correction. In Section~\ref{sec:4} we review a useful monogamy
relation satisfied by quantum and classical correlations in a
tripartite pure quantum state. Such a relation will be exploited in
Section~\ref{sec:5} to show that to have a small loss of entanglement
of formation is equivalent to have small classical correlations
between the reference system and the environment. This simple
observation will lead us to the main result stated as
Theorem~1. Section~\ref{sec:5a} extends the same analysis to other
entanglement measures. In particular, it is shown that for certain
entanglement measures it is possible to derive the same result as for
the entanglement of formation, but in a simpler way, moreover greatly
improving the tightness of the bound. This second result, independent
of the previous one, is stated as Theorem~2. Section~\ref{sec:disc}
stresses two remarks by comparing the two theorems obtained so
far. The first remark shows that they can be combined to explicitly
obtain the above mentioned inequality, regarding the gap between
entanglement of formation and distillable entanglement for a general
bipartite mixed state. The second remark proposes a possible
connection between different bipartite entanglement measures, used
here to derive different criteria for approximate quantum error
correction, and correspondingly induced topologies on the set of
quantum channels. A brief summary (Section~\ref{sec:summary})
concludes the paper.

\section{Tripartite purification of channels}\label{sec:2}

Let us consider an input quantum system $Q$ whose state is described
by the density matrix $\rho^Q$ defined on the (finite dimensional)
input Hilbert space $\sH^Q$. A \emph{channel}, mapping states on
$\sH^Q$ (that is, the set of nonnegative, trace-one operators on
$\sH^Q$, briefly denoted as $\states(\sH^Q)$) to states on $\sH^{Q'}$,
can be represented as a completely positive trace-preserving (CP-TP)
linear map $\mE:\states(\sH^Q)\to\states(\sH^{Q'})$. We will use the
notation $\rho^{Q'}:=\mE(\rho^Q)$. It is a well-known fact that
channels can be written in their so-called Kraus form~\cite{kraus},
that is
\begin{equation*}
  \mE(\rho^Q)=\sum_mE_m\rho^Q E_m^\dag,\qquad\forall\rho^Q,
\end{equation*}
where the Kraus operators $E_m$ satisfy the normalization condition
$\sum_mE_m^\dag E_m=\openone^Q$.

Besides the above mentioned abstract definition, we can give a
different description of channels, by exploiting a powerful
representation theorem, direct consequence of Stinespring
theorem~\cite{stine}, which states that all channels can be realized
by means of a suitable unitary interaction $U^{QE}$ of the input
system $Q$ with an \emph{ancilla} $E$ (initialized in a fixed pure
state $|0^E\>\in\sH^E$), followed by a trace over the ancillary
degrees of freedom, in formula
\begin{equation*}
  \mE(\rho^Q)=\Tr_{E'}\left[U^{QE}\ (\rho^Q\otimes|0\>\<0|^E)\ (U^{QE})^\dag\right].
\end{equation*}
(We put a prime also on $E$, because in general the output ancilla
system could be different from the input one.) Such a purification of
the channel can always be realized, without loss of generality, with
$\dim\sH^{E'}\le\dim\sH^Q \times\dim\sH^{Q'}$ and it is unique up to
local isometries on $\sH^{E'}$. Since in the following we will
consider entropic quantities, such an isometric freedom is completely
innocuous.

It is now convenient to introduce a third reference system $R$, which
purifies $\rho^Q$ as
\begin{equation*}
  \Psi^{RQ}:=|\Psi\>\<\Psi|^{RQ}\textrm{ such that }\Tr_R[\Psi^{RQ}]=\rho^Q.
\end{equation*}
As before, also this purification is unique up to local isometries on
$\sH^R$, so that $S(\rho^Q)=S(\rho^R)$, where
$\rho^R=\Tr_Q[\Psi^{RQ}]$ and $S(\sigma):=-\Tr[\sigma\log_2\sigma]$ is
the von Neumann entropy of the state $\sigma$. We can always choose,
without loss of generality, the reference to be isomorphic to the
input, so that $\dim\sH^R=\dim\sH^Q$. The reference system $R$ goes
untouched through the interaction $U^{QE}$, in such a way that the
\emph{global} state after the system-environment interaction is pure
and given by
\begin{equation}\label{eq:global}
  |\Psi^{RQ'E'}\>:=(\openone^R\otimes U^{QE})\ |\Psi^{RQ}\>\otimes|0^E\>.
\end{equation}
Since we closed the whole system, we will be able to play with
entropic quantities exploiting useful identities like
\begin{equation}\label{eq:useful}
  I^{R:Q'}(\rho^{RQ'})+I^{R:E'}(\rho^{RE'})=2S(\rho^{R})=2S(\rho^Q),
\end{equation}
where $I^{A:B}(\sigma^{AB}):=S(\sigma^A)+S(\sigma^B)-S(\sigma^{AB})$
is the quantum mutual information~\cite{strato,cerf} between $A$ and
$B$ when the global state is $\sigma^{AB}$, and $\rho^{RQ'}$
\emph{etc} are the reduced states calculated from the global
tripartite pure state $|\Psi^{RQ'E'}\>$ in Eq.~(\ref{eq:global}).

\section{Known conditions for channel correction}\label{sec:3}

How well does a channel $\mE$ preserve quantum information? That is,
how well does it preserve the entanglement that an unknown input state
shares with other systems? A way to give a quantitative answer to this
question is to introduce the \emph{entanglement fidelity}, that is a
nonnegative quantity, depending on the channel $\mE$ (we now suppose
that the output space coincides with the input one) and on the input
state $\rho^Q$, defined as~\cite{schum}
\begin{equation*}
  F(\rho^Q,\mE):=\<\Psi^{RQ}|(\id\otimes\mE)(\Psi^{RQ})|\Psi^{RQ}\>,
\end{equation*}
where $\Psi^{RQ}$ is a purification of $\rho^Q$ as before. It can be
proved that $F(\rho^Q,\mE)$ does not depend on the particular
purification $\Psi^{RQ}$ of $\rho^Q$, and it is an intrinsic property
of the channel, given the input state. If $F(\rho^Q,\mE)$ is close to
unity, then the channel $\mE$ acts almost like the identity channel
$\id$ on the support of $\rho^Q$, that is, every state in the support
of $\rho^Q$ is faithfully transmitted by $\mE$, along with its
eventual entanglement with other quantum systems.

Another quantity which tells how much a given channel preserves
coherence is given by the \emph{coherent information}
$\cohinfo(\rho^Q,\mE)$, defined as~\cite{schum-nielsen,lloyd}
\begin{equation*}
  \cohinfo(\rho^Q,\mE):=S(\rho^{Q'})-S(\rho^{RQ'})\le S(\rho^Q),
\end{equation*}
where, consistently with the notation introduced in the previous
section, $\rho^{Q'}:=\mE(\rho^Q)$ and
$\rho^{RQ'}:=(\id\otimes\mE)\Psi^{RQ}$. The coherent information can
be negative and it plays a fundamental role in quantifying the rate at
which a channel can reliably transmit quantum
information~\cite{lloyd,shor,devetak}.

Between entanglement fidelity and coherent information there exists a
close relation~\cite{schumacher-westmoreland} which states that, given
an input state $\rho^Q$ and a channel
$\mE:\states(\sH^Q)\to\states(\sH^{Q'})$, there exists a channel
$\mR:\states(\sH^{Q'})\to\states(\sH^Q)$ such that
\begin{equation}\label{eq:direct}
  F(\rho^Q,\mR\circ\mE)\ge 1-\sqrt{2\left(S(\rho^Q)-\cohinfo(\rho^Q,\mE)\right)}.
\end{equation}
In other words, if the coherent information is close to the input
entropy, then the channel can be approximately
corrected~\cite{nota2}. Most important, also the converse statement is
true, in the sense that a sort of quantum Fano inequality
holds~\cite{schum,barnischu}
\begin{equation}\label{eq:converse}
  S(\rho^Q)-\cohinfo(\rho^Q,\mE)\le\operatorname{g}(1-F(\rho^Q,\mR\circ\mE)),
\end{equation}
for all channels $\mR:\states(\sH^{Q'})\to\states(\sH^Q)$, where
$\operatorname{g}(x)$ is an appropriate positive, concave (and hence
continuous), monotonically increasing function such that $\lim_{x\to
  0}\operatorname{g}(x)=0$. In particular, for $x\le 1/2$, we can take
$\operatorname{g}(x):=4x\log_2(d/x)$, where
$d:=\dim\sH^Q$~\cite{schum,barnischu}. In other words, if a channel
$\mR$ happens to approximately correct the channel $\mE$, then
$\cohinfo(\rho^Q,\mE)$ has to be correspondingly close to the input
entropy. Notice that Eqs.~(\ref{eq:direct}) and~(\ref{eq:converse})
are nothing but entropic formulations of the fact that approximate
correction is possible if and only if the joint reference-ancilla
output state $\rho^{RE'}$ is close to being factorized, that is
$\rho^{RE'}\approx\rho^{R}\otimes\rho^{E'}$ (about this point, see
also Ref.~\cite{decoupling-approach}). In fact,
\begin{equation*}
\begin{split}
  S(\rho^Q)-\cohinfo(\rho^Q,\mE)&=I^{R:E'}(\rho^{RE'})\\
&=D(\rho^{RE'}\|\rho^{R}\otimes\rho^{E'}),
\end{split}
\end{equation*}
where $D(\rho\|\sigma):=\Tr[\rho\log_2\rho-\rho\log_2\sigma]$ is the
\emph{quantum relative entropy} and can be understood as a kind of
distance between states.

From Eqs.~(\ref{eq:direct}) and~(\ref{eq:converse}), it is an
immediate corollary that \emph{perfect} correction (on the support of
$\rho^Q$) is possible if and only if~\cite{schum-nielsen}
\begin{equation*}
  \cohinfo(\rho^Q,\mE)=S(\rho^Q).
\end{equation*}
However, coherent information is not the only quantity which enjoys
such a property. By introducing the \emph{entanglement of formation},
defined for a bipartite mixed state $\sigma^{AB}$ as~\cite{eof}
\begin{equation*}
\begin{split}
  \eof&(\sigma^{AB}):=\\
&\min_{\{p_i,|\phi_i^{AB}\>\}_i:\sum_ip_i\phi_i^{AB}=\sigma^{AB}}\sum_ip_iE\left(\phi_i^{AB}\right),
\end{split}
\end{equation*}
where the minimum is taken over all possible pure state ensemble
decomposition of $\sigma^{AB}$ as
$\sigma^{AB}=\sum_ip_i\phi_i^{AB}$ and
$E(\phi^{AB}):=S\left(\Tr_{B}\left[\phi^{AB}\right]\right)$ is
the \emph{entanglement} of the pure bipartite state $\phi^{AB}$, in
Ref.~\cite{schumacher-westmoreland-2} it is proved that \emph{perfect}
correction (on the support of $\rho^Q$) is possible if and only if
\begin{equation*}
\eof(\rho^{RQ'})=S(\rho^Q).
\end{equation*}
The ``only if'' part is not surprising, since it is known that (for an
elementary proof, see Section~\ref{sec:4} below)
\begin{equation}\label{eq:to-prove}
\cohinfo(\rho^Q,\mE)\le\eof(\rho^{RQ'}),
\end{equation}
and the above relation can hold \emph{strictly} (in fact, coherent
information can easily be negative). Hence we immediately obtain the
analogous of Eq.~(\ref{eq:converse})
\begin{equation}\label{eq:converse2}
  S(\rho^Q)-\eof(\rho^{RQ'})\le\operatorname{g}(1-F(\rho^Q,\mR\circ\mE)),
\end{equation}
that is, the existence of an approximately correcting channel $\mR$
implies that the entanglement of formation of $\rho^{RQ'}$ is close
to $S(\rho^Q)$,~\cite{nota1b}.

In Ref.~\cite{schumacher-westmoreland-2} it was left open the question
whether also the converse statement is true, namely if the
entanglement of formation of $\rho^{RQ'}$ is a \emph{robust} measure
of the correctability of a channel. Before answering (affirmatively)
this question, we have to go back to the unitary realization of
channels and give an alternative interpretation of the entanglement of
formation.

\section{Classical, quantum, and total correlations}\label{sec:4}

The entanglement of formation $\eof(\sigma^{AB})$ is a well-behaved
measure of the quantum correlations existing between two quantum
systems $A$ and $B$ described by the joint state $\sigma^{AB}$. On the
other hand, the quantum mutual information $I^{A:B}(\sigma^{AB})$
measures the \emph{total} correlations, quantum as well as classical,
that a bipartite quantum system exhibits~\cite{decorr}. Notice that
both entanglement of formation and quantum mutual information are by
construction symmetric under the exchange of $A$ and $B$.

On the contrary, the quantity measuring the amount of \emph{classical}
correlations in a bipartite quantum state loses such a symmetry, and a
logical direction of classical correlations seems to naturally
emerge. Such a quantity, proposed in Ref.~\cite{class-corr}, is
defined as
\begin{equation*}
\begin{split}
  C&^{B\to A}(\sigma^{AB}):=\\
&\max_{\{P_i^B\}_i}\left[S(\sigma^A)-\sum_ip_iS\left(\frac{\Tr_B\left[\sigma^{AB}\ \left(\openone^A\otimes P_i^B\right)\right]}{p_i}\right)\right],
\end{split}
\end{equation*}
where the maximum is taken over all possible POVMs $\{P_i^B\}_i$ (that
is, $P_i^B>0$ for all $i$, and $\sum_iP_i^B=\openone^B$) on the
subsystem $B$ and $p_i:=\Tr\left[\sigma^{B}P_i^B\right]$. Such
a measure is asymmetric, since in general $C^{B\to A}(\sigma^{AB})
\neq C^{A\to B}(\sigma^{AB})$, and it is closely related to the
assisted classical capacity of quantum channels~\cite{hayden-king}.

In Ref.~\cite{koashi-winter} it is proved that for a tripartite pure
state $|\phi^{ABC}\>$ the relation $C^{B\to A}(\sigma^{AB})+
\eof(\sigma^{AC})= S(\sigma^A)$ holds, where $\sigma^{AB}$ \emph{etc}
are the reduced states of $|\phi^{ABC}\>$. In the case of a channel,
given the global state $|\Psi^{RQ'E'}\>$ in Eq.~(\ref{eq:global}), we
correspondingly have
\begin{equation}\label{eq:monogamy}
\ccorr(\rho^{RE'})+\eof(\rho^{RQ'})=S(\rho^Q).
\end{equation}
We are now able to easily prove Eq.~(\ref{eq:to-prove}). In fact,
since $\cohinfo(\rho^Q,\mE)= S(\rho^Q)- I^{R:E'}(\rho^{RE'})$, and
from Eq.~(\ref{eq:monogamy}), thanks to the monotonicity of quantum
relative entropy under the action of channels, namely
$D(\rho\|\sigma)\ge D(\mE(\rho)\|\mE(\sigma))$,
$\forall(\rho,\sigma,\mE)$, we have that
\begin{equation}\label{eq:upperbound}
  \ccorr(\rho^{RE'})\le I^{R:E'}(\rho^{RE'}),
\end{equation}
which in turn directly implies
\begin{equation*}
  \cohinfo(\rho^Q,\mE)\le\eof(\rho^{RQ'}).
\end{equation*}

\section{Entanglement of formation and approximate channel
  correction}\label{sec:5}

In this section we will present the main result, that is, the loss of
entanglement of formation is small if and only if the channel can be
approximately corrected. We saw before that approximate correction is
possible if and only if the joint reference-ancilla output state
$\rho^{RE'}$ is almost factorized~\cite{schumacher-westmoreland}. We
would then like to say that the loss of entanglement of formation is
small if and only if $\rho^{RE'}$ is almost factorized.

The ``if'' part has already been written in the form of
Eq.~(\ref{eq:upperbound}). In fact, if
$\rho^{RE'}\approx\rho^{R}\otimes\rho^{E'}$, then
$S(\rho^{R}\otimes\rho^{E'})\approx S(\rho^{RE'})$ thanks to Fannes'
continuity property, which implies that $I^{R:E'}(\rho^{RE'})\approx
0$, and, in turn, that $\ccorr(\rho^{RE'})\approx 0$, or,
equivalently, that $\eof(\rho^{RQ'})\approx S(\rho^Q)$ (see
Eq.~(\ref{eq:monogamy})).

To prove the ``only if'' part is a little trickier. We exploit the
existence, proved in Ref.~\cite{infoc} for every dimension of
the Hilbert space, of (rank-one) \emph{informationally complete
  measurements}, that are POVMs whose elements form a basis for the
operator space. In other words, there always exists a POVM $\{P_i\}_i$
such that $\Tr[XP_i]=0$ for all $i$ if and only if $X=0$. Notice that
this is the generalization of the usual concept of quantum state
tomography.  Informationally complete POVMs have a (generally non
unique) dual set $\{\tilde P_i\}_i$ such that the following
reconstruction formula holds
\begin{equation}\label{eq:reconstr}
\sum_i\Tr[XP_i]\tilde P_i=X,\qquad\forall X.
\end{equation}
Notice that the dual operators $\tilde P_i$ are generally non
positive, but can always be chosen hermitian~\cite{postpro}. We are
now in position to write the following chain of inequalities
($\N{X}_1:= \Tr|X|$ denotes the trace-norm)
\begin{equation}\label{eq:chain}
\begin{split}
  \N{\rho^{RE'}-\rho^{R}\otimes\rho^{E'}}^2_1&=\N{\sum_ip_i\left(\rho^{R}_i\otimes\tilde P^{E'}_i-\rho^{R}\otimes\tilde P^{E'}_i\right)}_1^2\\
  &\le\sum_ip_i\N{\left(\rho^{R}_i-\rho^{R}\right)\otimes\tilde P^{E'}_i}_1^2\\
  &\le K\sum_ip_i\N{\rho^{R}_i-\rho^{R}}_1^2\\
  &\le 2K\sum_ip_iD(\rho^{R}_i\|\rho^{R})\\
  &\le 2K\ccorr(\rho^{RE'})\\
  &=2K\left(S(\rho^Q)-\eof(\rho^{RQ'})\right).
\end{split}
\end{equation}
Let us explain one by one all the passages in the above equation:
\begin{itemize}
\item[(i)] In the first line we applied identity~(\ref{eq:reconstr})
  to the subsystem $E'$, where $\{P_i^{E'}\}_i$ is an informationally
  complete POVM and $\{\tilde P_i^{E'}\}_i$ its dual frame, and
  defined $p_i:=\Tr[\rho^{E'}P_i^{E'}]$ and
  $\rho^{R}_i:=\Tr_{E'}[\rho^{RE'}\ (\openone^{R}\otimes
  P_i^{E'})]/p_i$.
\item[(ii)] In the second line we used the convexity of the function
  $x\mapsto x^2$.
\item[(iii)] In the third line we defined $K:=\max_i\N{\tilde
    P_i^{E'}}_1^2$, which is finite because we are considering finite
  dimensional Hilbert spaces.
\item[(iv)] In the fourth line we used Pinsker
  inequality~\cite{hayashi}, that is $\N{\rho-\sigma}^2_1\le
  2D(\rho\|\sigma)$.
\item[(v)] In the fifth line we simply used the fact that
  $\ccorr(\rho^{RE'})$ is defined as a \emph{maximum} over all
  possible measurements on $E'$.
\item[(vi)] In the last line we used Eq.~(\ref{eq:monogamy}).
\end{itemize}
Summarizing, we obtained that whenever $\ccorr(\rho^{RE'})\to 0$, or,
equivalently, $\eof(\rho^{RQ'})\to S(\rho^Q)$, then $\N{\rho^{RE'}-
  \rho^{R}\otimes \rho^{E'}}^2_1\to 0$ correspondingly, which in turn
implies the existence of an approximately correcting channel
$\mR$,~\cite{schumacher-westmoreland}. Notice that, as a trivial
corollary, we get that $C^{B\to A}(\sigma^{AB})=0$ if and only if
$\sigma^{AB}=\sigma^{A}\otimes\sigma^{B}$.

In the sequence of inequalities in Eq.~(\ref{eq:chain}), the most
unpleasant feature is the size of the constant $K$. In fact, it is
clearly independent of the channel and the input state, however, we
did not investigate how it depends on the dimensions of the input and
output Hilbert spaces $\sH^Q$ and $\sH^{Q'}$. We can give a rough
upper bound on $K$ by considering the (continuous outcome)
informationally complete POVM $\{P_g\}_{g\in\mathbb{SU}(d)}$ defined
as
\begin{equation*}
P_g:=\frac 1dU_g\varphi U_g^\dag,
\end{equation*}
where $U_g$ is a unitary representation of the group $\mathbb{SU}(d)$,
and $\varphi$ is a pure state. In Ref.~\cite{infoc} the canonical dual
set $\{\tilde P_g\}_g$ has been explicitly calculated, and it holds
that
\begin{equation*}
\N{\tilde P_g}_1=2d-1,\qquad\forall g,
\end{equation*}
where $d$ is the dimension of the Hilbert space on which the POVM
$\{P_g\}_g$ is measured, in our case $\sH^{E'}$. Since we saw that its
dimension can be upper-bounded as
$\dim\sH^{E'}\le\dim\sH^Q\times\dim\sH^{Q'}$, we obtain the following
\begin{equation}\label{eq:at_hand}
  \N{\rho^{RE'}-\rho^{R}\otimes\rho^{E'}}^2_1\le 2(2dd'-1)^2\left(S(\rho^Q)-\eof(\rho^{RQ'})\right),
\end{equation}
where $d:=\dim\sH^Q$ and $d':=\dim\sH^{Q'}$. Anyway, the only
assumption we need about the ancilla POVM $\{P_i^{E'}\}_i$ is that it
is informationally complete. We could hence use the one, among
informationally complete POVMs, whose dual set minimizes $K$. How to
choose such an ``optimal'' informationally complete measurement is
left as a wide open question.

At the end we can state the following:
\begin{theo}
  Given an input state $\rho^Q$, defined on the Hilbert space $\sH^Q$,
  and a channel $\mE$ mapping states on $\sH^Q$ to states on
  $\sH^{Q'}$, let us define $\varepsilon_\textrm{\emph f}:=S(\rho^Q)-
  E_\textrm{\emph f}(\rho^{RQ'})$. Then, there exists a channel $\mR$,
  from states on $\sH^{Q'}$ to states on $\sH^Q$, such that
\begin{equation}\label{eq:rhs}
F(\rho^Q,\mR\circ\mE)\ge 1-\sqrt{2(2dd'-1)^2\varepsilon_\textrm{\emph f}},
\end{equation}
where $d:=\dim\sH^Q$ and $d':=\dim\sH^{Q'}$.
\end{theo}

{\bf Proof.} With Eq.~(\ref{eq:at_hand}) at hand, the proof is
straightforward. It makes use of the well-known relation existing
between fidelity and trace-distance, that is
\begin{equation*}
\fid(\rho,\sigma)\ge 1-\frac{\N{\rho-\sigma}_1}{2},
\end{equation*}
and of the main result of Ref.~\cite{schumacher-westmoreland},
thanks to which the existence of a channel $\mR$ such that
\begin{equation*}
F(\rho^Q,\mR\circ\mE)\ge\fid^2(\rho^{RE'},\rho^{R}\otimes\rho^{E'})
\end{equation*}
is guaranteed. $\blacksquare$

\section{Other entanglement measures}\label{sec:5a}

Up to now, we considered the entanglement of formation $\eof$ as the
entanglement measure quantifying quantum correlations. Such a choice
is motivated by the fact that it is known~\cite{hayashi} that $\eof$
is an upper bound to the coherent information itself as well as to
many other genuine entanglement measures $E_\bullet$ (among these, for
example, one finds the \emph{distillable
  entanglement}~\cite{distillable}, the \emph{relative entropy of
  entanglement}~\cite{relative}, and the \emph{squashed
  entanglement}~\cite{squashed}, just to cite three of them). The
following corollary directly stems from Theorem~1
\begin{corollary}
  If $E_\bullet\le E_\textrm{\emph f}$, the following inequality holds
\begin{equation}\label{eq:generic_bound}
 F(\rho^Q,\mR\circ\mE)\ge 1-\sqrt{2(2dd'-1)^2\varepsilon_\bullet},
\end{equation}
where $\varepsilon_\bullet:=S(\rho^Q)-E_\bullet(\rho^{RQ'})$.
\end{corollary}
{\bf Proof.} Trivial. $\blacksquare$

Then, thanks to the above mentioned ``extremality property'' enjoyed
by the entanglement of formation among entanglement measures,
Corollary~1 can be applied to many different situations, making the
conclusions we drew form Theorem~1 quite general.

On the other hand, the so-called hashing inequality~\cite{hashing}
\begin{equation}\label{eq:hashing}
  \cohinfo(\rho^Q,\mE)\le E_\textrm{d}^{R\to Q'}(\rho^{RQ'})\qquad(\le E_\textrm{d}(\rho^{RQ'})),
\end{equation}
where $E_\textrm{d}(\sigma^{AB})$ is the distillable entanglement and
$E_\textrm{d}^{A\to B}(\sigma^{AB})$ is the \emph{one-way} distillable
entanglement (i.~e. we restrict the classical communication to go from
$A$ to $B$ only), implies the converse direction, namely, if $\ode\le
E_\bullet$, then the analogous of Eq.~(\ref{eq:converse}),
\begin{equation}\label{eq:lhs}
  S(\rho^Q)-E_\bullet(\rho^{RQ'})\le\operatorname{g}(1-F(\rho^Q,\mR\circ\mE)),
\end{equation}
holds true. It is worth stressing here that while the condition
$\ode\le E_\bullet$ is very general, the condition $E_\bullet\le \eof$
is satisfied by many among known entanglement measures but not by all
of them (a notable exception is, for example, the \emph{logarithmic
  negativity}~\cite{logneg}). Nevertheless, it is known that whatever
generic entanglement measure satisfying a certain number of conditions
can be proved to lie between $\ode$ and
$\eof$~\cite{christhesis}. Hence inequivalent entanglement measures,
provided they behave ``sufficiently well'', lead to equivalent
conditions for \emph{approximate} quantum error correction,
generalizing what was already noted in
Ref.~\cite{schumacher-westmoreland-2} in the case of \emph{exact}
correction.

By further specializing the entanglement measure, we can say more. If
the entanglement measure is chosen to be ``not too large'', it is
possible to refine the bound~(\ref{eq:generic_bound}) as
follows. More explicitly, the following result, that we state as a
second theorem independent from Theorem~1, can be proved
\begin{theo}
  Let $\mE$ be a channel acting on states on the input Hilbert space
  $\sH^Q$. Let $E_\bullet(\sigma^{AB})$ be an entanglement measure
  such that
\begin{equation}\label{eq:goodent}
E_\bullet(\sigma^{AB})\le\frac{I^{A:B}(\sigma^{AB})}{2},\qquad\forall\sigma^{AB}
\end{equation}
holds, and define $\varepsilon_\bullet:=S(\rho^Q)-
E_\bullet(\rho^{RQ'})$. Then, there exists a channel $\mR$ such that
\begin{equation}\label{eq:theo2}
F(\rho^Q,\mR\circ\mE)\ge 1-2\sqrt{\varepsilon_\bullet}.
\end{equation}
\end{theo}
{\bf Proof.} The proof goes as follows:
\begin{equation*}
\begin{split}
 \frac 12\N{\rho^{RE'}-\rho^{R}\otimes\rho^{E'}}^2_1&\le D(\rho^{RE'}\|\rho^{R}\otimes\rho^{E'})\\
&=2S(\rho^Q)-I^{R:Q'}(\rho^{RQ'})\\
&\le 2S(\rho^Q)-2E_\bullet(\rho^{RQ'}),
\end{split}
\end{equation*}
where we used again Pinsker inequality and Eq.~(\ref{eq:useful}). At
this point, by the same passages as in the proof of Theorem~1, we
obtain the statement. $\blacksquare$

Relation~(\ref{eq:theo2}) is clearly much tighter than the analogous
relation~(\ref{eq:generic_bound}), in that here we succeeded in
getting rid of the dependence on the dimensions of the input and
output Hilbert space. Notice that condition~(\ref{eq:goodent}) it is
proved to hold for the distillable entanglement and for the squashed
entanglement~\cite{squashed}. On the other hand, such a derivation
cannot be applied to the entanglement of formation which can be
smaller or larger than the quantum mutual entropy~\cite{generic_ent}.

\section{Relation between entanglement of formation and distillable
  entanglement}\label{sec:disc}

It is interesting to directly compare the three relations
(Eqs.~(\ref{eq:direct}),~(\ref{eq:generic_bound}),
and~(\ref{eq:theo2})) for approximate quantum error correction that we
considered throughout the paper:
\begin{equation}\label{eq:comparison}
\begin{split}
&F(\rho^Q,\mR\circ\mE)\ge 1-\sqrt{2(S(\rho^Q)-I_\textrm{c}(\rho^Q,\mE))},\\
&F(\rho^Q,\mR\circ\mE)\ge 1-2\sqrt{S(\rho^Q)-E_\bullet(\rho^{RQ'})},\\
&F(\rho^Q,\mR\circ\mE)\ge 1-\sqrt{2(2dd'-1)^2(S(\rho^Q)-E_\bullet(\rho^{RQ'}))},
\end{split}
\end{equation}
where $d:=\dim\sH^Q$ and $d':=\dim\sH^{Q'}$. The first is proved in
Ref.~\cite{schumacher-westmoreland}, the second holds if
$E_\bullet(\sigma^{AB})\le I^{A:B}(\sigma^{AB})/2$, while the third
holds if $E_\bullet(\sigma^{AB})\le E_\textrm{f}(\sigma^{AB})$. The
numerical factor in front of the ``loss figure'' gets larger as we
move from coherent-information--loss toward
entanglement-of-formation--loss. This feature is reminiscent of the
fact that, in general, the gap $\eof> E_\textrm{d}$ between
entanglement of formation and distillable entanglement can be
generically large~\cite{generic_ent}.

Concerning this point, it is interesting to notice that our approach
can be somehow useful to understand to which extent such a gap can be
authentically arbitrary. In fact, entanglement of formation and
distillable entanglement coincide on pure states, and both of them are
known to be asymptotically continuous in the mixed neighborhood of
every pure state~\cite{christhesis}. It is then reasonable that,
sufficiently close to pure states, entanglement of formation and
distillable entanglement become equivalent entanglement measures (in
the sense that they can be reciprocally bounded), and the gap between
them cannot be completely arbitrary. In fact we can say something
more in the form of the following
\begin{corollary}
  For an arbitrary bipartite mixed state $\sigma^{AB}$, with
  $S(\sigma^A)\le S(\sigma^B)$, let us define the coherent information
\begin{equation*}
I_\textrm{\emph c}^{A\to B}(\sigma^{AB}):=S(\sigma^B)-S(\sigma^{AB})
\end{equation*}
and the entanglement of formation deficit
\begin{equation*}
\varepsilon_{\textrm{\emph
    f}}(\sigma^{AB}):=S(\sigma^A)-E_\textrm{\emph f}(\sigma^{AB}).
\end{equation*}
Then, the following inequality holds
\begin{equation}\label{eq:gap}
S(\sigma^A)-I_\textrm{\emph c}^{A\to B}(\sigma^{AB})
\le\operatorname{g}\left(\sqrt{2(2d_Ad_B-1)^2\varepsilon_\textrm{\emph
      f}(\sigma^{AB})}\right),
\end{equation}
where $\operatorname{g}(x)$ is a function as in
Eq.~(\ref{eq:converse}), and $d_{A(B)}:=\dim\sH^{A(B)}$.
\end{corollary}
{\bf Proof.} First of all, let us notice that whatever
bipartite mixed state $\sigma^{AB}$ can be written as
$(\id^A\otimes\mE^B)(\Psi^{AB})$, for some channel $\mE^B$ and some
pure $\Psi^{AB}$ such that $\Tr_{B}[\Psi^{AB}]=\sigma^A$. This simple
observation is in order to make sure that all equations, previously
obtained for bipartite states
$\rho^{RQ'}=(\id^R\otimes\mE^Q)(\Psi^{RQ})$, can be in particular
interpreted as equations valid for all bipartite mixed states
$\sigma^{AB}$ as well, simply paying attention to the directionality
intrinsic in the definition of coherent information. Then, putting
together Eqs.~(\ref{eq:converse}) and~(\ref{eq:rhs}), we obtain the
statement~(\ref{eq:gap}). $\blacksquare$

The large numerical factor multiplying $\varepsilon_\textrm{f}$ in
Eq.~(\ref{eq:gap}) makes it possible the above mentioned generic gap
exhibited by high dimensional systems, that is, entanglement of
formation can be close to the maximum value, while distillable
entanglement is null (or almost null). Eq.~(\ref{eq:gap}) then says
``how large'', for fixed finite dimensions $d_A$ and $d_B$, the gap
can actually be: in fact we can affirm that if the entanglement of
formation is ``sufficiently close'' to its maximum value, then also
the coherent information and, thanks to the hashing
inequality~(\ref{eq:hashing}), the one-way distillable entanglement
have to be ``correspondingly large''. Notice moreover that there may
be room for a further improvement of Eq.~(\ref{eq:gap}), since we
obtained it as coming from a probably over-simplified estimation. To
tighten the evaluation of the constant $K$ in Eq.~(\ref{eq:chain})
could then be useful in understanding the relationships between
entanglement of formation and distillable entanglement as well,
besides being an interesting mathematical problem by itself.

Before concluding, we would like to stress one more remark. It is
clear from Eq.~(\ref{eq:comparison}) how we are actually dealing with
three different topologies on the set of quantum channels induced by
different measures of bipartite entanglement~\cite{private}. Also this
connection definitely deserves further investigation.

\section{Conclusions}\label{sec:summary}

In summary, we generalized the information-theoretical analysis of
approximate quantum error correction based on coherent information
given in Ref.~\cite{schumacher-westmoreland}, by showing that
approximate quantum error correction is possible if and only if the
loss of entanglement along the quantum channel is small. We considered
explicitly different entanglement measures, in particular the
entanglement of formation and the distillable entanglement, showing
how equivalent conclusions come from inequivalent entanglement
measures. We moreover showed that the approach used here can be
applied also to understand the interconnections existing between
entanglement of formation and distillable entanglement, even though
they are known to behave quite independently, in particular in high
dimensional quantum systems.

\appendix

\acknowledgements The author acknowledges Japan Science and
Technology Agency for support through the ERATO-SORST Quantum
Computation and Information Project. Thank you to Masahito Hayashi and
Lorenzo Maccone for useful comments and suggestions.


\begin{thebibliography}{000}
\bibitem{nota1} The literature about the subject is huge and rapidly
  growing. For a reasonably recent and compact review of seminal
  papers see Ref.~\cite{fault-tol}.
\bibitem{fault-tol} D~Gottesman, in {\small \it Encyclopedia of
    Mathematical Physics}, eds. J-P~Fran\c{c}oise, G~L~Naber and
  S~T~Tsou, (Elsevier, Oxford, 2006), vol.~4, pp.~196-201. Available
  online as arXiv:quant-ph/0507174v1.
\bibitem{nielsen} M~A~Nielsen and I~L~Chuang, {\small \it Quantum
    Computation and Quantum Information} (Cambridge University Press,
  Cambridge, 2000), pp.~425-499.
\bibitem{algebraic-exact} J~Kempe, in {\small \it Quantum
    Decoherence, Poincar\'e seminar 2005}, Progress in Mathematical
  Physics Series, (Birkhaeuser Verlag, 2006), p.~85-123. Available
  online as arXiv:quant-ph/0612185v1.
\bibitem{schum-nielsen} B~Schumacher and M~A~Nielsen,
  Phys.~Rev.~A~{\bf 54}, 2629 (1996).
\bibitem{ogawa} T~Ogawa, arXiv:quant-ph/0505167v2.
\bibitem{nielsen-poul} M~A~Nielsen and D~Poulin,
  arXiv:quant-ph/0506069v1.
\bibitem{schumacher-westmoreland} B~Schumacher and M~D~Westmoreland,
  Quant.~Inf.~Processing~{\bf 1}, 5 (2002).
\bibitem{schumacher-westmoreland-2} B~Schumacher and M~D~Westmoreland,
  J.~Math.~Phys.~{\bf 43}, 4279 (2002).
\bibitem{generic_ent} P~Hayden, D~W~Leung, and A~Winter,
Comm.~Math.~Phys.~{\bf 265}, 95 (2006).
\bibitem{kraus} K~Kraus, {\small \it States, Effects, and Operations:
    Fundamental Notions in Quantum Theory}, Lect.~Notes~Phys.~{\bf
    190}, (Springer-Verlag, Berlin, 1983).
\bibitem{stine} W~F~Stinespring, Proc.~Am.~Math.~Soc.~{\bf 6}, 211
  (1955).
\bibitem{strato} R~L~Stratonovich, Prob.~Inf.~Transm.~{\bf 2}, 35
  (1965).
\bibitem{cerf} C~Adami and N~J~Cerf, Phys.~Rev.~A~{\bf 56}, 3470
  (1997).
\bibitem{schum} B~Schumacher, Phys.~Rev.~A~{\bf 54}, 2614 (1996).
\bibitem{lloyd} S~Lloyd, Phys.~Rev.~A~{\bf 55}, 1613 (1997).
\bibitem{shor} P~W~Shor, ``The quantum channel capacity and coherent
  in- formation,'' Lecture Notes, MSRI Workshop on Quantum
  Computation, San Francisco, 2002 (unpublished). Available online at
  http://www.msri.org/publications/ln/msri/2002/\\quantumcrypto/shor/1
\bibitem{devetak} I~Devetak, IEEE Trans.~Inf.~Theory~{\bf 51}, 44
  (2005).
\bibitem{nota2} In Ref.~\cite{schumacher-westmoreland} the following
  inequality is discussed
\begin{equation*}
F(\rho^Q,\mR\circ\mE)\ge 1-2\sqrt{\left(S(\rho^Q)-\cohinfo(\rho^Q,\mE)\right)},
\end{equation*}
which is indeed a little looser than Eq.~(\ref{eq:direct}). It is
however clear, already from the arguments used there, that
Eq.~(\ref{eq:direct}) actually holds true.
\bibitem{barnischu} H~Barnum, M~A~Nielsen, and B~Schumacher,
  Phys.~Rev.~A~{\bf 57}, 4153 (1998).
\bibitem{decoupling-approach} P~Hayden, M~Horodecki, J~Yard, and
  A~Winter, arXiv:quant-ph/0702005v1.
\bibitem{eof} C~H~Bennett, D~P~Di~Vincenzo, J~A~Smolin, and
  W~K~Wootters, Phys.~Rev.~A~{\bf 54}, 3824 (1996).
\bibitem{nota1b} In fact, $\eof(\sigma^{AB})\le
  \min\{S(\sigma^A),S(\sigma^B)\}$, $\forall\sigma^{AB}$, holds, so
  that l.~h.~s. of Eq.~(\ref{eq:converse2}) is positive.
\bibitem{decorr} B~Groisman, S~Popescu, and A~Winter,
  Phys.~Rev.~A~{\bf 72}, 032317 (2005).
\bibitem{class-corr} L~Henderson and V~Vedral, J.~Phys.~A:
  Math.~Gen.~{\bf 34}, 6899 (2001).
\bibitem{hayden-king} P~Hayden and C~King (2005),
  Quantum~Inform.~Comput.~{\bf 5}, 156 (2005).
\bibitem{koashi-winter} M~Koashi and A~Winter, Phys.~Rev.~A~{\bf 69},
  022309 (2004).
\bibitem{infoc} G~M~D'Ariano, P~Perinotti, and M~F~Sacchi, J.~Opt.~B:
  Quantum and Semicl.~Optics~{\bf 6}, S487 (2004).
\bibitem{postpro} G~M~D'Ariano and P~Perinotti, Phys.~Rev.~Lett.~{\bf
    98}, 020403 (2007).
\bibitem{hayashi} M~Hayashi, {\small \it Quantum Information: an
    Introduction} (Springer-Verlag, Berlin, Heidelberg, 2006).
\bibitem{distillable} C~H~Bennett, H~J~Bernstein, S~Popescu, and
  B~Schumacher, Phys.~Rev.~A~{\bf 53}, 2046 (1996).
\bibitem{relative} V~Vedral, M~B~Plenio, M~A~Rippin, and P~L~Knight,
  Phys.~Rev.~Lett.~{\bf 78}, 2275 (1997).
\bibitem{squashed} M~Christandl and A~Winter, J.~Math.~Phys.~{\bf 45},
  829 (2004).
\bibitem{hashing} I~Devetak and A~Winter, Proc.~Roy.~Soc.~London
  A~{\bf 461}, 207 (2004).
\bibitem{logneg} G~Vidal and R~F~Werner, Phys.~Rev.~A~{\bf 65}, 032314
  (2002).
\bibitem{christhesis} M~Christandl, arXiv:quant-ph/0604183v1.
\bibitem{private} M~Hayashi, private communication.
\end{thebibliography}
\end{document}